\documentclass[applemac]{aa}
\usepackage{inputenc}
\usepackage{natbib}
\usepackage{txfonts}
\usepackage{graphicx}
\usepackage{color}
\bibpunct{(}{)}{;}{a}{}{,}

\begin{document}

\title{On the stability of nonisothermal Bonnor-Ebert spheres. III. The role of chemistry in core stabilization}
\author{O. Sipil\"a\inst{1},
		P. Caselli\inst{1},
	     \and{M. Juvela\inst{2}}
}
\institute{Max-Planck-Institute for Extraterrestrial Physics (MPE), Giessenbachstr. 1, 85748 Garching, Germany \\
e-mail: \texttt{osipila@mpe.mpg.de}
\and{Department of Physics, P.O. Box 64, 00014 University of Helsinki, Finland}
}

\date{Received / Accepted}

\authorrunning{O. Sipil\"a et al.}

\abstract
{}
{We investigate the effect of chemistry on the stability of starless cores against gravitational collapse.}
{We combine chemical and radiative transfer simulations in the context of a modified Bonnor-Ebert sphere to model the effect of chemistry on the gas temperature, and study the effect of temperature changes on core stability.}
{We find that chemistry has in general very little effect on the nondimensional radius $\xi_{\rm out}$ which parametrizes the core stability. Cores that are initially stable or unstable tend to stay near their initial states, in terms of stability (i.e., $\xi_{\rm out} \sim$ constant), as the chemistry develops. This result is independent of the initial conditions. We can however find solutions where $\xi_{\rm out}$ decreases at late times ($t \gtrsim 10^6\,\rm yr$) which correspond to increased stabilization caused by the chemistry. Even though the core stability is unchanged by the chemistry in most of the models considered here, we cannot rule out the possibility that a core can evolve from an unstable to a stable state owing to chemical evolution. The reverse case, where an initially stable core becomes ultimately unstable, seems highly unlikely.
}
{Our results indicate that chemistry should be properly accounted for in studies of star-forming regions, and that further investigations of core stability especially with hydrodynamical models are warranted.}


\maketitle

\section{Introduction}

The Bonnor-Ebert sphere (hereafter BES; \citealt{Bonnor56}; \citealt{Ebert55}), essentially an isothermal gas sphere in hydrostatic equilbrium, has been used in many studies to represent the physical structure of starless/prestellar cores, perhaps most famously in the case of B68 by \citet{Alves01}. It is however an observationally established fact that real cores are not isothermal \citep{Crapsi07, Pagani07}. BES models including the possibility of non-isothermality have been studied by several authors \citep{Evans01, Zucconi01, Galli02, Keto05, Juvela11}. \citet{Keto08, Keto10} and \citet{Keto14, Keto15} have studied the particular case of L1544 using hydrodynamics coupled with (simplified) chemistry. In the previous studies considering static cloud models, the dust or gas temperature was either calculated based on the density profile of a BES so that the model retained the physical structure of the isothermal core, or the equations determining the density profile were modified to include the possibility of a radially-varying temperature. We refer to a core model constructed using the latter approach as a modified Bonnor-Ebert sphere (MBES).

When attempting to reproduce observations using a particular core model, it is necessary to consider whether the model core is stable. The stability condition of the BES is well known \citep{Bonnor56}: cores with nondimensional radius $\xi_{\rm out} \gtrsim 6.45$ represent unstable configurations. The stability of a BES with internal motions or subjected to non-ionizing radiation has been studied by \citet{Seo13} and \citet{Seo16}. However, the stability condition of the MBES is a relatively little-studied subject. In \citet{Sipila11} (hereafter Paper~1), we presented an analytical expression for the stability condition of the MBES based on an extension of Bonnor's equations to accommodate radial changes in the temperature. In that paper we made the assumption that $T_{\rm dust} = T_{\rm gas}$, which is approximately valid only at high density. In \citet{Sipila15c} (Paper~2), we expanded the validity range of the stability analysis by switching from $T_{\rm dust}$ to $T_{\rm gas}$ in the relevant equations, made possible by including chemistry in the calculation framework. In both papers it was found that the stability condition of the MBES, parametrized by the non-dimensional radius $\xi_{\rm out}$, is close to that of the BES.

In this paper, we investigate the potential impact that chemistry may have on the stability of an evolving MBES, i.e., instead of studying the global effects on $\xi_{\rm out}$ in cores with varying masses as was done in our previous papers, we concentrate on the evolution of $\xi_{\rm out}$ for single cores. We select cores with varying initial nondimensional radii (stable and unstable configurations based on the results of Papers 1 and 2), and study the evolution of the physical properties of the core in tandem with the chemical evolution of the gas. The particular aim of the present paper is to find out if a core can evolve from the unstable regime into the stable regime, or vice versa, owing to chemical evolution. This important question has received little attention in the literature so far.

The paper is organized as follows. In Sect.\,\ref{s:model} we introduce our model and explain how the stability calculations are carried out in practice. The results of our analysis are presented in Sect.\,\ref{s:results}. In Sect.\,\ref{s:discussion} we discuss our results and in Sect.\,\ref{s:conclusions} we give our conclusions.

\section{Model}\label{s:model}

\subsection{Basic formulae}

The properties of the MBES are discussed in detail in Papers~1 and 2. Here, we recall the relevant formulae. The density distribution of the MBES is given by
\begin{equation}\label{eqrhodist}
\frac{1}{r^2} \frac{d}{dr} \left( \frac{r^2}{\rho} \left[ T \frac{d\rho}{dr} + \rho \frac{dT}{dr} \right] \right) = - \frac{4\pi Gm \rho}{k} \, ,
\end{equation}
which assumes the ideal gas equation of state and hydrostatic equilibrium. In the above, $m$ and $k$ denote the average molecular mass\footnote{Even though in this paper we employ a detailed chemical model that includes a wide range of chemical species, we make the approximation in the calculation of the molecular mass that the gas consists of $\rm H_2$ and He only so that $m = 2.33$\,amu.} of the gas and the Boltzmann constant, respectively, and $\rho$ denotes the medium density. By making the substitutions
\begin{equation}\label{eqrho}
\rho = \frac{\lambda}{\tau(\xi)} {\rm e}^{-\psi(\xi)}
\end{equation}
\begin{equation}\label{eqr}
r = \beta^{1/2}\lambda^{-1/2}\xi \, ,
\end{equation}
one can transform Eq.\,(\ref{eqrhodist}) into nondimensional form:
\begin{equation}\label{eqMLE}
\xi^{-2} \frac{d}{d\xi} \left[ \xi^2 \tau \frac{d\psi}{d\xi} \right] = \frac{1}{\tau} {\rm e}^{-\psi} \, ,
\end{equation}
the modified Lane-Emden equation. In the above, $\xi$ represents the nondimensional radius of the core ($\xi$ and $\psi(\xi)$ are both dimensionless); $\lambda$ represents the central density; $\tau(\xi) = T(\xi)/T_{\rm c}$ represents the radial (gas) temperature profile.

The variable $\beta$ in Eq.\,(\ref{eqr}) represents the (scaled) velocity dispersion at the center of the core. In Papers~1 and 2 we assumed that the dispersion is due to purely thermal motions, and hence that $\beta = \sigma_{\rm s, th}^2 / 4\pi G =  k T_{\rm c} / 4\pi G m$. Here, we introduce a turbulent component to the velocity dispersion with a constant value of 200\,$\rm m\,s^{-1}$, which is in the range of typical values observed in dense cores \citep[e.g.,][]{Pattle15}. Therefore in this paper
\begin{equation}\label{beta}
\beta = \frac{\sigma_{\rm s, th}^2 + \sigma_{\rm s, nt}^2}{4\pi G} \, ,
\end{equation}
where $\sigma_{\rm s, th}^2 = k T_{\rm c} / m$ represents the thermal component and $\sigma_{\rm s, nt} = 200\,\rm m\,s^{-1}$ represents the non-thermal component. We discuss the effect of assuming a purely thermal velocity distribution on our results in Sect.\,\ref{ss:noturb}.

Equation (\ref{eqMLE}) can be solved numerically when the temperature profile is known. Here, we determine the dust and gas temperature profiles by using radiative transfer calculations (see below). The following boundary conditions are imposed at the core center: $\psi = 0$, $d\psi / d\xi = 0$, $\tau = 1$, and $d\tau / d\xi = 0$, which ensure that $T = T_{\rm c}$ and $\rho = \lambda$ at $\xi = 0$, i.e., at the center of the core.

\subsection{Method}

In Paper~2, we examined the stability of the MBES under the assumption that the non-dimensional radius $\xi_{\rm out}$ and the total mass $M$ of a model core stay constant as the gas temperature changes with the chemical evolution of the gas. As discussed in Paper~2, these assumptions may not hold in reality, which means that the results presented in that paper represent the possible configurations that MBESs of various masses can exist in, when embedded in regions of different chemical ages. Here, we change our approach and keep the external pressure $p_{\rm ext}$ constant so that the model represents a core evolving inside a fixed environment. The boundary pressure can be written as
\begin{equation}\label{pressure}
p_{\rm ext} = \rho_{\rm out} \left( \sigma_{\rm s, th}^2 + \sigma_{\rm s, nt}^2 \right) = \frac{4 \pi G \beta \lambda e^{-\psi_{\rm out}}}{\tau_{\rm out}},
\end{equation}
where $\tau(\xi) = T(\xi) / T_{\rm c}$ (see Paper~1) and the subscript ``out'' refers to a value taken at the nondimensional outer radius of the core, $\xi_{\rm out}$, which is here a time-dependent variable contrary to the scenario discussed in Papers~1 and 2. Note that the expression of $p_{\rm ext}$ in the above is different than the one given in Paper~1 (Eq.\,4 therein) because here the $\beta$ parameter also includes a non-thermal contribution.

In addition to the external pressure, we are free to fix one of two properties of the core: the central density or the mass (while $\xi$ is not fixed). Here we choose to fix the central density. This choice allows us to determine the nondimensional radius $\xi_{\rm out}$ using Eq.\,(\ref{pressure}). By tracking the changes in $\xi_{\rm out}$ as the gas temperature changes with the chemical evolution of the core, we can study the potential impact of the chemistry on the core stability. We explain below how this is achieved in practice.

Setting the core mass and the external pressure as the problem constants leads sometimes to situations where a solution cannot be found, i.e., the evolution of $\xi_{\rm out}$ cannot be determined. We discuss models with fixed mass in more detail in Appendix~\ref{appendix_a}.

\subsection{Determining the core evolution}\label{ss:coreev}

\begin{figure}
\centering
\includegraphics[width=\columnwidth]{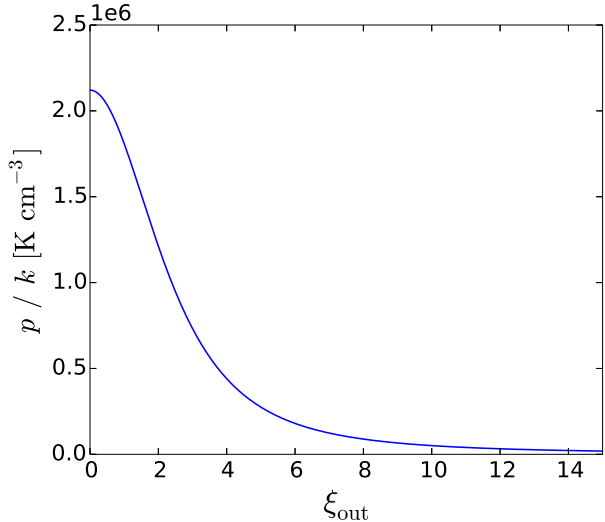}
\caption{Boundary pressure of a BES with central density $\lambda = 10^5 \, \rm cm^{-3}$, $T = 10 \, \rm K$ as a function of nondimensional radius $\xi_{\rm out}$.
}
\label{fig:BESpext}
\end{figure}

We constructed the initial MBES by first constructing a BES with predetermined values of central density, (gas and dust) temperature and external pressure. The choice of the external pressure is however not completely free. This is demonstrated by Fig.\,\ref{fig:BESpext}, where we plot the boundary pressure of a BES as a function of $\xi_{\rm out}$. Here the central density is set to $\lambda = 10^5 \, \rm cm^{-3}$ and the temperature (constant throughout the core) to $T = 10 \, \rm K$; we also assumed a turbulent component corresponding to a velocity dispersion of $\sigma_{\rm s, nt} = 200\, \rm m\,s^{-1}$. Different values of $\xi_{\rm out}$ correspond to different configurations that a core with the chosen values of central density and temperature can exist in. The steepness of the density gradient within a core increases with $\xi_{\rm out}$. Also, the core mass, for example, is different for each value of $\xi_{\rm out}$. The value of the boundary pressure for the configuration with $\xi_{\rm out} = 0$ represents the maximum value of $p_{\rm ext}$ possible for a BES with the chosen parameters. The lower the boundary pressure, the more unstable the core configuration is, since solutions with $\xi_{\rm out} \gtrsim 6.5$ represent unstable cores \citep{Bonnor56}.

Here, when constructing the initial MBES, we chose the boundary pressure of the BES to be lower than the maximum value allowed by the choices of central density and initial central temperature. We then calculated the dust temperature of the BES using the \citet{OH94} dust model (see also Papers~1 and 2). The initial MBES was constructed in the following steps: 1) solve Eq.\,(\ref{eqMLE});  2) find the value of $\xi$ that satisfies Eq.\,(\ref{pressure}); 3) calculate the density profile using Eq.\,(\ref{eqrho}). After the initial MBES was set up, we proceeded to calculate the chemical evolution in the core following the methods laid out in Paper~2; we separated the model core into spherical shells and calculated the chemical evolution separately in each shell. The initial conditions of the model are discussed in Sect.\,\ref{ss:inicond}. The abundances of various cooling species were extracted from the chemical modeling results and used as input for the radiative transfer code LOC (Juvela, in prep.) to determine the gas temperature profile, which was in turn used to update the MBES following the steps described above. The gas temperature is determined by the relative strengths of heating and cooling processes. The gas is assumed to be heated by cosmic rays, and cools by line radiation and loss of energy in collisions with dust grains. In this paper we used the gas-dust coupling scheme of \citet{Goldsmith01}. We note that a stronger coupling between gas and dust \citep[e.g.,][]{Young04} would reduce the role of chemistry in determining the gas temperature because it increases the range of physical conditions where the gas-dust coupling is important.

We assumed in our models that the visual extinction at the edge of the core is $A_{\rm V} = 2 \, \rm mag$. However, the radiative transfer calculations also include a layer outside the core itself which absorbs line radiation, and can emit it back into the core, preventing the gas temperature from dropping to very low values (6-7\,K) at the edge of the core at early times when the abundances of the main coolants are high. The external molecular layer is assumed to have the same density as the edge of the core, and its thickness is set so to correspond to one magnitude of visual extinction. The layer is not extended all the way to $A_{\rm V} = 0 \, \rm mag$ because there $\rm H_2$ should dissociate, and in this work we only desire to model what happens inside the core. However, the inclusion or exclusion of the external molecular layer, or the assumed amount of extinction at the core edge, have only a very limited effect on our results (see Sects.\,\ref{s:results}~and~\ref{ss:extprop}). We note that in this paper we do not model possible temporal changes in dust properties. The unattenuated ISRF spectrum is taken from \citet{Black94}.

We adopt the chemical model from \citet{Sipila15a} and \citet{Sipila15b}, although in this paper we do not include deuterium in the chemical calculations in order to speed up the process. We adopt the initial chemical abundances from Table~3 in \citet{Sipila15a}, where the gas is initially atomic with the exception of hydrogen which is in $\rm H_2$. We discuss the effect of changing the initial abundances in Sect.\,\ref{ss:inicond}.

In contrast to Paper 2, we consider here a time-dependent approach to the chemistry. Starting from the initial MBES ($t = t_0$), we solve the chemistry until some predetermined time step $t_1$, and after solving the gas temperature and updating the density profile of the core following the steps 1 to 3 discussed above, we pick up the chemistry at $t_1$ and continue until $t_2$, and so on. In this way we can track how the nondimensional radius $\xi_{\rm out}$ evolves as a function of time, and study whether the stability of the core is affected, using also the results from Papers~1~and~2. We studied four combinations of central densities and boundary pressures (see Table~\ref{tab1}), allowing chemical development up to $10^7$\,yr. We set the length of the initial time step to 100 years, and considered 19 subsequent time steps logarithmically spaced between $10^2$ and $10^7$\,yr. Tests varying the amount of time steps were carried out and it was found that our main results, described below, remained virtually unaffected by changes in the time resolution.

The values of the core parameters ($\lambda$, $p_{\rm ext}$) were chosen in such a way that we obtained a range of core masses and nondimensional radii (see below). We concentrated on intermediate-density cores to bring out the effect of the chemistry; at high density the dust and gas are efficiently coupled through collisions and the role of the chemistry in determining the gas temperature diminishes.

\begin{table}
\caption{Central densities ($\lambda$) and boundary pressures ($P_{\rm ext} / k$) of the core models discussed in this paper.}
\centering
\begin{tabular}{c | c | c}
\hline \hline 
Core & $\lambda \, [\rm cm^{-3}]$ & $P_{\rm ext} / k \, [\rm K \, cm^{-3}]$ \\ \hline
A & $1.0 \times 10^5$ & $1.0 \times 10^5 $ \\
B & $1.0 \times 10^5$ & $5.0 \times 10^5 $ \\
C & $5.0 \times 10^4$ & $7.0 \times 10^4 $ \\
D & $5.0 \times 10^4$ & $3.0 \times 10^5 $ \\
\hline
\end{tabular}
\tablefoot{$\lambda$ represents the $\rm H_2$ density, $n({\rm H_2})$.}
\label{tab1}
\end{table}

\section{Results}\label{s:results}

\begin{figure}
\centering
\includegraphics[width=\columnwidth]{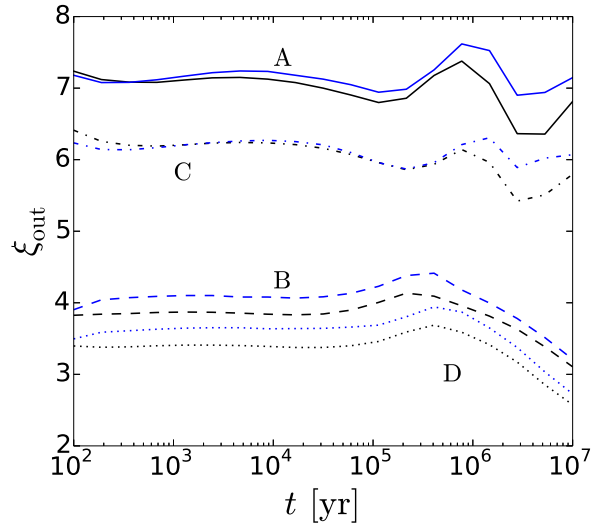}
\caption{Nondimensional radii $\xi_{\rm out}$ of the four cores studied in this paper as functions of time. The core models (see Table~1) are indicated in the figure. The black lines represent our fiducial model described in the text, while the blue lines represent a model where the external molecular layer outside the core is not included.
}
\label{fig:MBES_xiout}
\end{figure}

\begin{figure*}
\centering
\includegraphics[width=2.0\columnwidth]{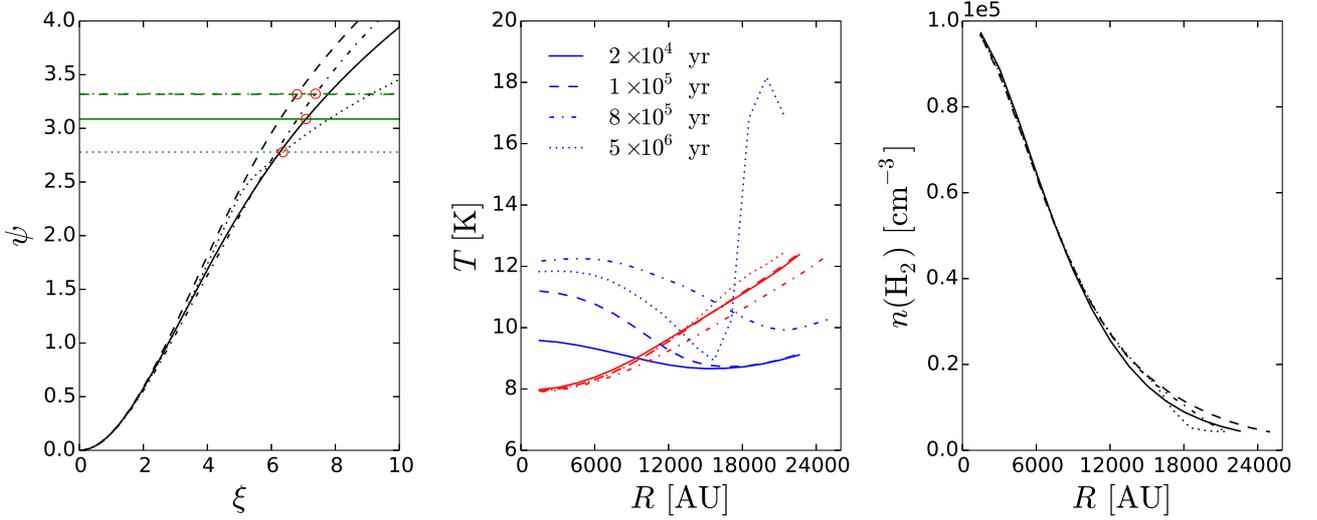}
\caption{{\sl Left-hand panel:} function $\psi$ for core A at selected time steps, indicated in the middle panel of the figure. The green horizontal lines mark the value of $\psi$ that satisfies Eq.\,(\ref{pressure}) at each time step, and the red circles are a guide for the eye to locate the corresponding value of $\xi_{\rm out}$ (see text). {\sl Middle panel:} dust (red) and gas (blue) temperature profiles for core A as functions of distance from the core center at different times. {\sl Right-hand panel:} density profile for core A as a function of distance from the core center at different times. The quoted times are approximate because the time steps considered in the model, save for the first and last, are not exact multiples of ten.
}
\label{fig:MBES_props}
\end{figure*}

\begin{figure*}
\centering
\includegraphics[width=2.0\columnwidth]{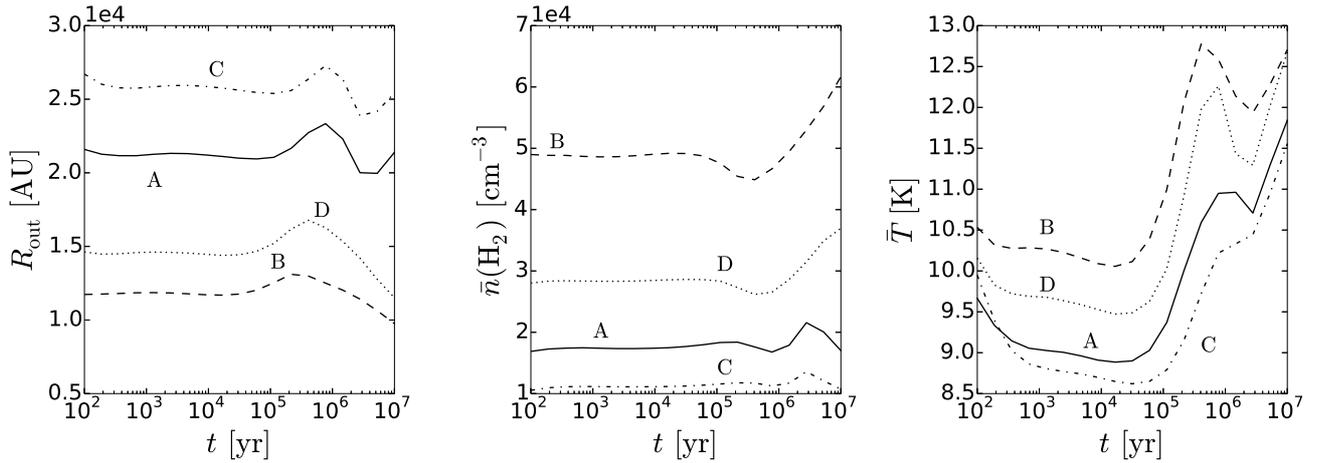}
\caption{Outer radii (left-hand panel), average densities (middle panel), and average temperatures (right-hand panel) of the four cores studied in this paper (indicated in the figure) as functions of time.
}
\label{fig:MBES_averages}
\end{figure*}

We plot in Fig.\,\ref{fig:MBES_xiout} the evolution of $\xi_{\rm out}$ as a function of time for the four core models tabulated in Table~\ref{tab1}. A clear trend is seen in all of the cores: $\xi_{\rm out}$ tends to stay approximately constant until $t \sim 10^5 \, \rm yr$. In two of the cores (B and D) $\xi_{\rm out}$ begins to decrease at $t > 10^5 \, \rm yr$, while cores A and C present somewhat more complex behavior. The evolution of $\xi_{\rm out}$ can be understood by studying the evolution of the function~$\psi$ (which is related to the density profile; see Eq.\,\ref{eqrho}). Any changes in $\psi$ are tied to the evolution of the temperature profile. As an example, we show in Fig.\,\ref{fig:MBES_props} the function $\psi$, the temperature profile, and the density profile for core~A for selected time steps in the model. At early times the high abundance of CO helps to maintain a gas temperature of less than 10\,K at the center of the core, but as CO depletes the gas temperature is determined solely by the gas-dust coupling which results in a temperature difference of $\sim$4\,K between the two components at a density of $\sim 10^5 \, \rm cm^{-3}$ \citep[see also][]{Goldsmith01}. At late times the central temperature of the core changes very little as a function of time (i.e., $\beta \sim \rm constant$), and so the value of $\xi_{\rm out}$ is controlled mainly by $\psi_{\rm out}$ and $\tau_{\rm out}$ (see Eq.\,\ref{pressure}), with the former being determined by the temperature profile through the Lane-Emden equation (Eq.\,\ref{eqMLE}). We show in the left-hand panel of Fig.\,\ref{fig:MBES_props} the solution of Eq.\,(\ref{pressure}), i.e., $\psi_{\rm out} = - \log (p_{\rm ext} \tau_{\rm out} / 4\pi G \beta \lambda)$ for different time steps; the value of $\xi_{\rm out}$ at each time step is found at the intersection of a horizontal line with a $\psi$ curve of the same linestyle.

Note that the gas temperature rises sharply toward the core edge at $t \gtrsim 10^6\,\rm yr$. This effect is a result of the depletion of the cooling species onto grain surfaces, which takes a long time at the low density of the core edge. Then, most of the carbon is locked in grain-surface methane which is not readily broken into C or CO by photodissociation, depriving the gas of the main coolant species.

To complement the results discussed above, we show in Fig.\,\ref{fig:MBES_averages} the outer radii, average densities and average temperatures (as calculated with Eq.\,(B.4) in Paper~1) of the four cores studied here. The evolution of the outer radius follows that of $\xi_{\rm out}$. Cores B and D, which show an eventual decrease of $\xi_{\rm out}$, display contraction and an associated increase in average density at late times. The evolution of the average temperature is similar for all cores, showing an increase of 2-3\,K from early times to late times owing to molecular freeze-out. However, cores B and D are warmer than cores A and C because they start at lower values of $\xi_{\rm out}$ and hence their density contrast (central density vs. edge density) is lower. The higher densities toward the edge translate on the one hand to stronger molecular freeze-out and on the other hand to stronger gas-dust coupling. Both effects lead to increased average gas temperatures with respect to the cores with the same central density but higher density contrast.

Figure~\ref{fig:MBES_xiout} also shows the results of calculations where the external molecular layer is not included in the modeling. This change decreases the gas temperature at the core edge because of the increased photon escape probability. However, the two approaches to the details of the radiative transfer modeling produce the same general conclusions on the evolution of the core stability as a function of time. In particular, this shows that the temporal variation in $\xi_{\rm out}$ at late times is a real effect of the model and is not induced by our underlying assumptions.

\begin{figure}
\centering
\includegraphics[width=\columnwidth]{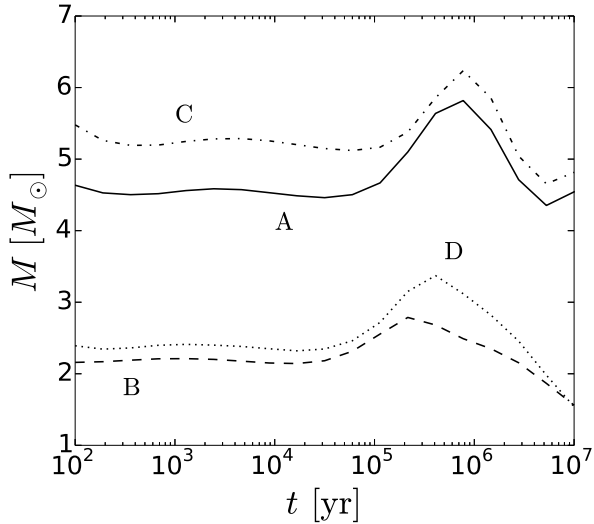}
\caption{Masses of the four cores studied in this paper (indicated in the figure) as functions of time.
}
\label{fig:MBES_mass}
\end{figure}

Because we fixed the boundary pressure and central density, the core mass is not a constant. The mass is given by (see Papers~1 and 2)
\begin{equation}\label{eq_mass_lambda}
M = 4\pi \, \beta^{3/2} \lambda^{-1/2} \xi_{\rm out}^2 \, \tau_{\rm out} \, \biggl(\frac{d\psi}{d\xi}\biggr)_{\rm out} \, .
\end{equation}
We plot in Fig.\,\ref{fig:MBES_mass} the masses of the four cores considered in this paper as functions of time. Evidently, the evolution of the core mass is similar to that of $\xi_{\rm out}$. This is expected based on Eq.\,(\ref{eq_mass_lambda}): the central density is a constant of the model and the central temperature varies generally very little as a function of time (at least at late times, see Fig.\,\ref{fig:MBES_props}), so that the change in mass is governed mainly by $\xi_{\rm out}^2 \, \tau_{\rm out} \, (\frac{d\psi}{d\xi})_{\rm out}$ which is dominated by $\xi_{\rm out}^2$. The changes in core mass imply that the core exchanges mass with its environment in order to meet the condition of fixed central density while the outer radius evolves.

The results presented here show that the nondimensional radius $\xi_{\rm out}$ tends to stay approximately constant with time -- or to decrease at late times -- in spite of chemical evolution within the core. Thus, chemical evolution may make no difference to the stability of the core, or may in some cases make it less likely for a core to collapse gravitationally. We discuss this issue further in Sect.\,\ref{ss:evo}.

\section{Discussion}\label{s:discussion}

\subsection{Effect of the external medium}\label{ss:extprop}

The present analysis is based on the assumption that the central density and the boundary pressure remain constant during core evolution. From a physical standpoint, considering the central density as a constant is reasonable if we aim to model a cloud core in quiescent conditions where little dynamical evolution should take place. However, setting the boundary pressure to a constant value means that we also constrain the properties of the medium outside the core. As mentioned in Sect.\,\ref{ss:coreev}, we assumed in the above analysis that the (gas) temperature of the external medium is equal to whatever value applies at the core boundary at any given time step. This implies also that the medium density outside the core is equal to that at the core edge, assuming that the turbulent velocity component has the same constant value in the core and in the surrounding cloud. Here we aim to study the temporal evolution of $\xi_{\rm out}$ in a general setting and we do not attempt to fit our results to any observed core. We simply assume that the properties of the external medium are at all times such that our model assumptions are fulfilled.

\begin{figure}
\centering
\includegraphics[width=\columnwidth]{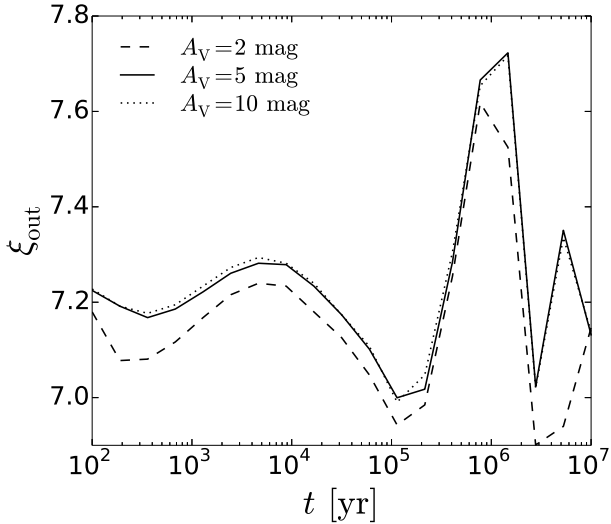}
\caption{Nondimensional radius $\xi_{\rm out}$ of core A as a function of time, assuming different values of $A_{\rm V}$ at the core edge.
}
\label{fig:MBES_xiout_diffav}
\end{figure}

However, for the sake of consistency with Papers 1~and~2, we studied the effect of varying the assumed visual extinction at the core edge on our results. The extinction (assumed equal to 2 mag in the results presented above) is expected to influence on our results, because it determines the strength of the external heating of the core and thus contributes to the gas temperature profile. The external heating in the present model is provided by the ISRF \citep{Black94}, and the photoelectric effect \citep[see][]{Juvela03b, Juvela11}. Increasing the extinction will decrease the gas temperature especially near the core edge, affecting the solution to Eq.\,(\ref{eqMLE}) and thus the derived value of $\xi_{\rm out}$. To quantify the possible effect on our results, we ran the core~A model with three different external $A_{\rm V}$ values (2, 5 and 10 mag), corresponding to situations where the model core lies progressively deeper inside a parent cloud. In these calculations the external molecular layer was not included. The results of this test are shown in Fig.\,\ref{fig:MBES_xiout_diffav}. Evidently, increasing the extinction increases $\xi_{\rm out}$, but the difference between the test cases is very small, in fact clearly smaller than the difference arising from including or excluding the external molecular layer (Fig.\,\ref{fig:MBES_xiout}).

\subsection{Initial conditions}\label{ss:inicond}

The analysis presented above is based on the assumption that the density and temperature structures of the cores correspond to modified BESs, i.e., that a (dust) temperature gradient exists already in the beginning of the core evolution. Furthermore, we assumed that the gas is initially atomic with the exception of hydrogen which is entirely in molecular form. In this section we explore the effect of changing these initial assumptions on our results.

Although CO is usually the main gas coolant \citep{Goldsmith01}, cooling by atomic C can also be important at early times \citep{Sipila12}. Therefore, it is not immediately clear if our results would be greatly affected if we assumed that (some) CO exists already in the initial state of the core, which is equivalent to the reasonable assumption that CO production has taken place preceding the formation of the core itself. To test this hypothesis we repeated the stability calculations presented above, now assuming initial abundances which were determined by a single-point chemical model assuming $n({\rm H_2}) = 10^4 \, \rm cm^{-3}$, $T = 12 \, \rm K$, and $A_{\rm V} = 2 \, \rm mag$. This model roughly represents the conditions in a molecular cloud and is here referred to as the ``molecular'' model. We started from an initially atomic chemical state and let the intermediate-density cloud evolve for $t = 10^5 \, \rm yr$, and used the resulting abundances as initial abundances in the stability calculation.

The initial abundances for the species relevant to the cooling calculations are given in Table~\ref{tab2}. Because of the low visual extinction ($A_{\rm V} = 2 \, \rm mag$), CO is efficiently destroyed by photons and its abundance stays at a constant level from $\sim 10^3 \, \rm yr$ to a few $\times 10^5 \, \rm yr$, never exceeding that of atomic carbon, which also stays more or less constant in abundance during this time interval. Therefore the time at which the abundances are extracted is of little consequence here. For higher values of external $A_{\rm V}$, translating to less efficient photoprocesses, CO would be the main carbon reservoir at this density.

\begin{figure}
\centering
\includegraphics[width=\columnwidth]{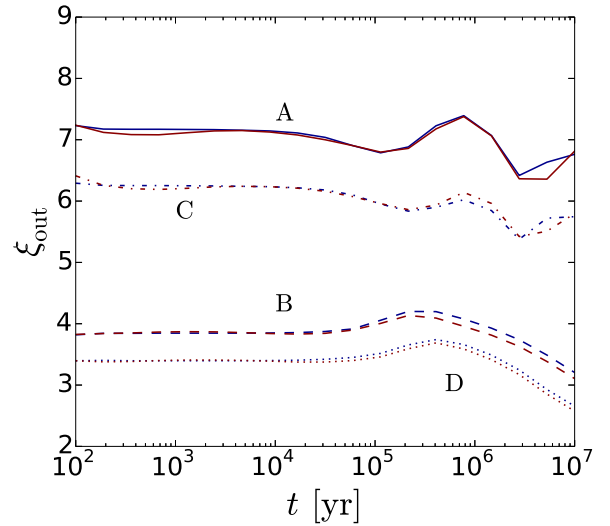}
\caption{Nondimensional radii $\xi_{\rm out}$ of the four cores studied in this paper as functions of time. Linestyles are the same as those in Fig.\,\ref{fig:MBES_xiout}. Red lines reproduce the data from Fig.\,\ref{fig:MBES_xiout}, while blue lines correspond to calculations in the molecular model (see text).
}
\label{fig:MBES_xiout_molini}
\end{figure}

Fig.\,\ref{fig:MBES_xiout_molini} shows the nondimensional radii $\xi_{\rm out}$ as functions of time in the molecular model. The results are very close to those presented in Fig.\,\ref{fig:MBES_xiout}; the $\xi_{\rm out}$ values of each core hardly change when molecular initial abundances are considered. The cooling power of C (in our fiducial model) and CO (in the initially molecular model) is similar at early times so that the model results do not deviate from each other much; all of the other relevant quantities ($\psi$, mass etc.) are also only marginally affected by the choice of the initial abundances. We emphasize that changes in the initial abundances do not cause a late-time increase of $\xi_{\rm out}$, which either stays approximately constant or decreases eventually.

\begin{table}
\caption{Initial abundances of species relevant to the cooling calculations.}
\centering
\begin{tabular}{c | c | c}
\hline \hline 
Species & Atomic & Molecular \\ \hline
CO & $0.0$ & $1.76 \times 10^{-5}$ \\
C$^+$,C & $1.20 \times 10^{-4}$ & $5.59 \times 10^{-5}$ \\
O & $2.56 \times 10^{-4}$ & $1.46 \times 10^{-4}$ \\
\hline
\end{tabular}
\tablefoot{``Atomic'' stands for the initial abundances adopted from Table~3 in \citet{Sipila15a}, where carbon is initially ionized. ``Molecular'' stands for abundances extracted from an intermediate-density cloud after $10^5 \, \rm yr$ of chemical evolution where carbon is mostly in atomic form (see text).
}
\label{tab2}
\end{table}

We also investigated the possibility that the initial core is isothermal, i.e., corresponds to a BES. We found that this assumption also makes very little difference to our results; as long as we choose the same boundary pressure and central density in both cases, starting from a BES or an MBES yields the same $\xi_{\rm out}$ values within $\sim$1\,\% after $t \sim 10^2$\,yr, i.e., the first time step in the model. Already at the first time step the difference between the models is less than $\sim$10\,\%. Our main results thus seem robust given their weak dependence on the initial conditions.

\subsection{Models without turbulent pressure}\label{ss:noturb}

\begin{figure}
\centering
\includegraphics[width=\columnwidth]{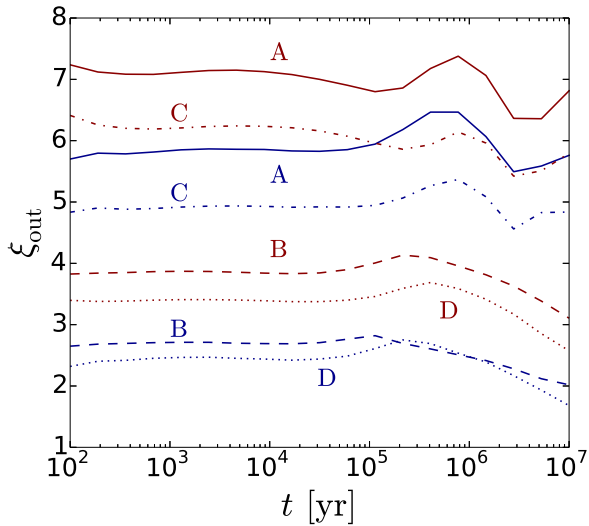}
\caption{Nondimensional radii $\xi_{\rm out}$ of the four cores studied in this paper as functions of time, calculated with a model including thermal velocity dispersion only. Linestyles are the same as those in Fig.\,\ref{fig:MBES_xiout}. Red lines reproduce the data from Fig.\,\ref{fig:MBES_xiout}, while blue lines correspond to calculations in the molecular model (see text).
}
\label{fig:MBES_xiout_nont}
\end{figure}

For completeness, we ran another set of calculations eliminating the nonthermal component from Eqs.\,(\ref{beta}) and (\ref{pressure}) to study whether this has a fundamental effect on our results. Figure \ref{fig:MBES_xiout_nont} shows the results of these calculations. It is observed that the $\xi_{\rm out}$ values are generally scaled down with respect to the model that includes turbulent support. This result is expected because removing the turbulent component reduces the maximum boundary pressure allowed for a given combination of central density and temperature, and thus the boundary pressures used (Table~\ref{tab1}) correspond to lower values of $\xi_{\rm out}$ in a model where turbulent pressure is neglected (see also the discussion in Sect.\,\ref{ss:coreev}). Our general conclusions are however the same in both types of model; $\xi_{\rm out}$ tends to stay constant or decrease at late times, and does not increase ultimately.

\subsection{Possible core evolution towards the stable regime}\label{ss:evo}

The lifetimes of molecular clouds and cloud cores are not strictly constrained. Recently, \citet{Brunken14} derived an age of at least $10^6$\,yr for the molecular envelope surrounding the protostar IRAS 16293-2422 using the spin states of $\rm H_2D^+$ as a chemical clock. Other estimates for the ages of molecular clouds range from $\sim$$10^6$\,yr \citep{Hartmann01} to up to $\sim$10\,Myr \citep{Mouschovias06}. The age of a molecular cloud sets a natural upper limit to the ages of cloud cores residing within it. Lifetimes of up to $\sim$$10^6$\,yr have been derived also for the cores themselves \citep{Ward-Thompson94, Kong16}. Given that the free-fall collapse timescale is $\sim$$10^5$\,yr for typical starless core densities, it seems clear that stabilizing forces must be at play for the cores/clouds to live up to Myr timescales. Magnetic fields may also play a significant role in the dynamics of molecular clouds by providing pressure to act as a means of support \citep[for observational and theoretical discussions see, e.g.,][]{Mouschovias06, Pillai15, Li15}.

Our results indicate that significant stabilization as a result of chemical evolution is not expected. Of the four cores studied here, cores B~and~D show decreasing values of $\xi_{\rm out}$ at late times, but these cores are stable to begin with. Cores A~and~C, which are initially unstable, show oscillating behavior at late times, but do not deviate significantly from their initial states in terms of stability. Based on our results one cannot rule out the possibility of a core configuration that starts out unstable and evolves into a stable state. However, our results do imply that such a transition would likely occur in a timescale of $\sim 10^6$\,yr, which agrees with the upper end of the range of expected cloud lifetimes estimated with theoretical and observational arguments. The fact that in general the chemistry has little to no effect on the stability of the cores is in line with recent calculations by \citet{Hocuk16}, where the consideration of gas-grain chemistry was not found to affect star formation rates, although their study was limited to the cloud scale and did not probe core evolution.

Finally we note that our models represent equilibrium solutions, and it is implicitly assumed that the dynamical timescales are shorter than the chemical timescales so that the cores are able to physically adapt to the changes in the chemistry. This type of an idealized situtation may not occur in real cores. To investigate the role of chemistry on core stability in a more realistic setting, the chemical and radiative transfer models should be combined with a (magneto)hydrodynamical model of core evolution. A similar study has been undertaken previously by, e.g., \citet{Keto05}. However, this type of investigation is beyond the scope of the present paper, and is reserved for future work.

\section{Conclusions}\label{s:conclusions}

We studied the stability of modified Bonnor-Ebert spheres with a time-dependent model that combines chemical and radiative transfer calculations for a consistent determination of the gas temperature as a function of time. The stability of the model cores was determined based on the results of our previous papers \citep{Sipila11, Sipila15c}. Unlike in our previous works which were based on the conservation of core mass and non-dimensional radius $\xi_{\rm out}$, we assumed here that the central density and external pressure are constant which describes the evolution of a core in a fixed quiescent environment. We assumed that the internal velocity dispersion is composed of a thermal and a nonthermal component.

We found that the non-dimensional radius $\xi_{\rm out}$ tends to remain approximately constant as a function of time, or in some cases to decrease late into the evolution of the core. An increase in instability, at least in a reasonable timescale, seems highly unlikely based on our models. However, the most likely scenario is that the chemistry does not make a significant difference to the core stability. Our fiducial model starts with an inially atomic gas (with the exception of hydrogen which is in $\rm H_2$). Changing the initial conditions was found to influence the values of $\xi_{\rm out}$ that we derive, but does not change our conclusions of the evolution of core stability.

We cannot completely rule out the possibility that chemistry can in certain conditions play a role in determining the lifetime of a starless core, particularly in the case where the core is on the border of being gravitationally stable or unstable. Further studies of the interplay between chemistry and core stability are called for, and may shed more light on the (low-mass) star-forming process in general. Here we considered equilibrium solutions with the assumption that the dynamical timescales are shorter than the chemical timescales. More comprehensive studies of the connection between chemistry and core stability should incorporate a description of dynamics in the form of (magneto)hydrodynamic modeling.

\begin{acknowledgements}
We thank the anonymous referee for useful comments and suggestions that helped to improve the manuscript. P.C. acknowledges financial support of the European Research Council (ERC; project PALs 320620). M.J. acknowledges the support of the Academy of Finland Grant No. 285769.
\end{acknowledgements}

\bibliographystyle{aa}
\bibliography{30146.bib}

\appendix

\section{Models with fixed mass and external pressure}\label{appendix_a}

As an alternative to the method discussed in the main part of this paper, where we fix the central density and boundary pressure of the core, we could instead fix the core mass and external pressure. In this case we eliminate $\lambda$ from Eqs.\,(\ref{pressure}) and (\ref{eq_mass_lambda}), which yields the relation
\begin{equation}\label{eq_fixedmass}
\xi^2 \, \left(\frac{d\psi}{d\xi}\right) \, \tau^{1/2} \, e^{-\psi/2} = \frac{M}{(4\pi)^{3/2}} \, \beta^{-2} \, G^{-1/2} \, p_{\rm ext}^{1/2} \, .
\end{equation}
The new value of $\xi_{\rm out}$ is determined by the function on the left, while the right side of the equation is just a number. The most meaningful comparison between the two approaches (fixed central density vs. fixed mass) is achieved by starting from the same initial condition and finding out if the evolution of $\xi_{\rm out}$ proceeds differently in the two cases. We tested this alternative method for the four cores studied in the paper (Table~\ref{tab1}).

\begin{figure}
\centering
\includegraphics[width=\columnwidth]{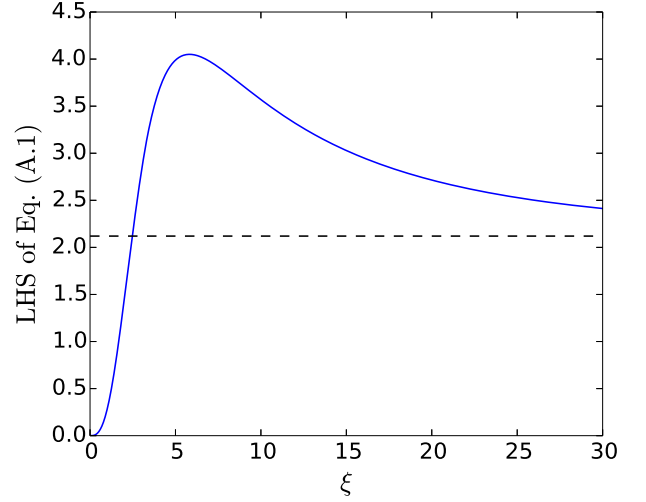}
\caption{The left-hand side (LHS) of Eq.\,(\ref{eq_fixedmass}) as a function of $\xi$ for the first time step for the core~B model assuming fixed core mass. The dashed line marks the value of the right-hand side of Eq.\,(\ref{eq_fixedmass}).
}
\label{fig:fixmp_mbes}
\end{figure}

\begin{figure}
\centering
\includegraphics[width=\columnwidth]{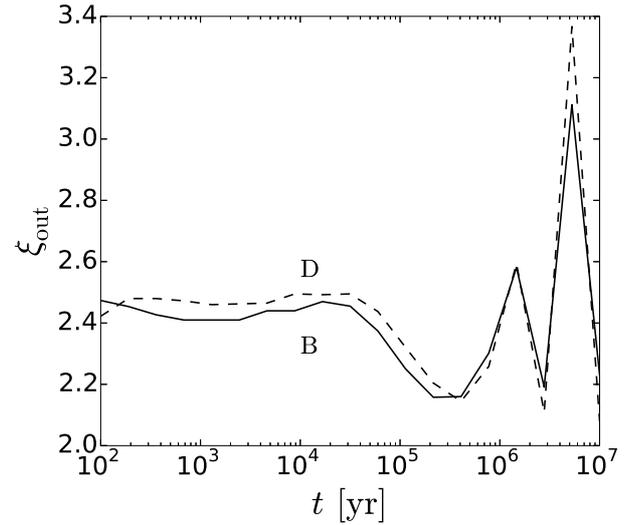}
\caption{Nondimensional radii $\xi_{\rm out}$ of cores B and D (indicated in the figure) as functions of time assuming fixed core mass.
}
\label{fig:MBES_xiout_fpm}
\end{figure}

The initial state of core~B corresponds to $M \sim 1.3 \, M_{\odot}$ and $\xi_{\rm out} \sim 3.4$. We proceeded with the same workflow as for the fixed central density case, i.e., we calculated the chemistry and the gas temperature followed by a solution of the Lane-Emden equation. This gives the shape for the left side of Eq.\,(\ref{eq_fixedmass}) shown in Fig.\,\ref{fig:fixmp_mbes}. One solution exists at $\xi_{\rm out} \sim 2.5$, which is now the new nondimensional radius of the core. Continuing the iterative process leads to the evolution of $\xi_{\rm out}$ shown in Fig.\,\ref{fig:MBES_xiout_fpm}, where we also show the evolution of $\xi_{\rm out}$ for core~D. The late-time fluctuation in $\xi_{\rm out}$ for both cores corresponds to an oscillation of the central density by a factor of $\sim$3. The timescale of the oscillation is affected by the chosen time resolution.

The fixed mass method does not work in all cases. For core~A the situtation at the first time step is such that no solution to Eq.\,(\ref{eq_fixedmass}) can be found. For core~C the same situation occurs at the second time step. The fixed mass method is more sensitive to the shape of the temperature profile, and to how the temperature profile continues beyond the core boundary, than the fixed central density method. Therefore the former is more ambiguous. It is conceivable that solutions could be found also for cores~A~and~C if different model assumptions were adopted.

\end{document}